\documentclass[aps,prd,preprint,showpacs]{revtex4}
\usepackage{latexsym}
\usepackage{amssymb}
\usepackage{graphicx}
\begin{document}

\thispagestyle{empty}

{\baselineskip0pt
\leftline{\large\baselineskip16pt\sl\vbox to0pt{\hbox{\it Department of
Mathematics and Physics}
               \hbox{\it Osaka City  University}\vss}}
\rightline{\large\baselineskip16pt\rm\vbox to20pt{\hbox{OCU-PHYS-306}
            \hbox{AP-GR-63}
\vss}}%
}

\vskip0.5cm

\title{High-Speed Collapse of a Hollow Sphere of Type I Matter}
\author{$^1$Zahid Ahmad\footnote{zahid$\_$rp@yahoo.com}, 
$^2$Tomohiro Harada\footnote{harada@rikkyo.ac.jp}, 
$^3$Ken-ichi Nakao\footnote{knakao@sci.osaka-cu.ac.jp}
and $^1$M. Sharif\footnote{msharif@math.pu.edu.pk}
}
\affiliation{$^1$Department of Mathematics, University of the Punjab, 
Quaid-e-Azam Campus, Lahore-54590, Pakistan\\
$^2$Department of Physics, Rikkyo University, Toshima, Tokyo 171-8501,~Japan\\
$^3$Department of Mathematics and Physics, Graduate School of Science, 
Osaka City University, Osaka 558-8585, Japan
}
\date{\today}

\begin{abstract}                
In this paper, we study the dynamics of a hollow 
spherical matter collapsing with very large initial velocity. The spacetime 
is initially very similar to the Vaidya solution, and the deviations 
from this background are treated perturbatively. 
The equations of state for radial pressure $p_{\rm R}=k\rho$ 
and tangential one $p_{\rm T}=w\rho$ with constant $k$ and $w$ are assumed.   
We find for the case of equations of state $k< 1$ and $0<w\leq1$ 
that the initial velocity, which is nearly the speed of light, is strongly 
decelerated. This result implies that the pressure is essential to the property 
of singularity formation in gravitational collapse even for 
initially nearly light-speed collapse. 
By contrast, in cases with the negative tangential pressure, 
the present result implies that the central naked singularity similar to 
that of the Vaidya spacetime 
can be formed, even though the radial pressure is positive, and 
the weak, strong and dominant energy conditions hold. 
Especially, in the case of $w<-(1-k)/4$, the high-speed collapse 
will produce the spacetime structure very similar to that of the Vaidya 
spacetime. 

\end{abstract}


\maketitle

\section{Introduction}
Einstein presented the theory of general relativity which
describes the gravitational force in terms of the spacetime
curvature. He formulated the field equations which relate the
geometry of the spacetime to matter fields. The earliest exact solutions
of these equations were the Schwarzschild metric representing the
exterior of a spherically symmetric star and the Friedmann
cosmological models. These solutions have spacetime singularities 
where the energy density or spacetime curvature diverges
and the usual description of the spacetime is impossible there\cite{R1}.

It is a well known phenomenon that gravitational collapse of massive objects
results in the formation of spacetime singularities in our
universe. There are two kinds of spacetime singularities from a point 
of view of their visibilities. A spacetime singularity is said to be
\emph{naked} when it is observable to local or distant observers, and the 
remaining one is said to be \emph{covered}. The singularity theorems
of Hawking and Penrose show that the formation of 
spacetime singularities is not rare in our universe if 
the general relativity is correct and the matter or radiation 
fields satisfy physically reasonable energy conditions\cite{R2}. 
However, these theorems do not provide
information about the visibility of the spacetime singularity.

About the visibility of the singularity, Penrose proposed
a conjecture called the cosmic censorship conjecture which 
has two versions\cite{R3}. The weak version states 
that the spacetime singularities produced by gravitational
collapse of physically reasonable matter fields, which develops 
from generic non-singular 
initial data, are always covered by horizon, whereas the strong version 
claims that there is no singularity visible to any observers. 
Some rather serious counterexamples have been found for the strong 
version\cite{R3-1,R3-2}.  As for the weak version, 
any precise theoretical or mathematical proof has not yet been 
given, although it has many
physical applications in black hole and other areas in
astrophysics. This motivates that a detailed study of dynamically
developing gravitational collapse models is necessary to obtain a
correct form of the cosmic censorship. Several examples (\cite{R4}-\cite{R12}
and references therein) have been studied so far which admit both
black hole and naked singularity solutions depending on the choice
of the initial data. Most of the work on gravitational collapse
has been done by considering dust fluid due to the existence of an
exact solution. However, the assumption of dust fluid might be too restricted as
the effects of pressure can not always be neglected in the formation 
processes of the spacetime singularities. 
Thus it is important to discuss this issue by including pressure.

Ori and Piran \cite{R13}-\cite{R15} investigated self-similar
spherically symmetric perfect fluid collapse by assuming the
equation of state $p=k\rho$. They found that a naked singularity
is formed for $0<k\lesssim0.0105$. They have also shown that there
exist naked-singular solutions with oscillations in the velocity
field for $0<k\leq0.4$. Later, these results were extended for
$0<k\leq0.5625$ by Foglizzo and Henriksen \cite{R16}. The same
results were also provided without self-similarity assumption by
one of the present author TH and Maeda\cite{R17}-\cite{R19}. 
Giambo et al. have
investigated naked singularity formation in perfect fluid collapse
without self-similarity assumption analytically\cite{R20}.
Goswami and Joshi \cite{R21} have also investigated analytically 
the local geometry near the central shell focusing singularity formed 
by a spherical collapse of a perfect fluid with the equation of 
state in the form $p=k\rho$ and have shown that the initial condition 
is crucial for whether this singularity is naked. However, 
it is difficult to construct a general global solution analytically. 
Towards the progress of analytical studies, 
it is important to develop a new approximation scheme (analytical
procedure) to discuss gravitational collapse which leads more
definite results.

One of the present authors KN and Morisawa studied  
the cylindrically symmetric gravitational collapse
of a thick shell composed of dust by introducing high-speed
approximation scheme\cite{R22}. The same authors generalized this
work for the perfect fluid case\cite{R23}. In a recent paper \cite{R24}, two of the 
present authors MS and ZA have extended this work by considering two perfect fluids. 
These investigations have provided interesting results about the
gravitational collapse. It would be worthwhile to explore whether
these results hold for spherical collapse or not. This motivated
us to develop a high-speed approximation scheme for spherically
symmetric system with the type I matter\cite{HE}. 
There are many studies about the naked singularity formation 
by the same type of matter in the spherically symmetric system. 
Dwivedi and Joshi showed by the local analysis that the initial data 
is crucial 
for whether the naked singularity forms at the symmetric
center by the gravitational collapse of general type I matter\cite{DJ1994}.
Recently, this work extended to the higher dimensional 
spacetime by Goswami and Joshi\cite{GJ2007}. The spherically 
symmetric matter with vanishing radial pressure and non-vanishing 
tangential pressure has been studied 
by various authors\cite{Magli1998,MGJM2002,TIN1998,TNI1999,GJ2002,MHJN2007}. 
In this paper, we consider the case treated by Ref.\cite{DJ1994} but 
we construct global analytic solutions by using 
high-speed approximation. 

The paper is organized as follows. In section II, we
write down the Einstein equations for the spherically symmetric spacetime 
with a type I matter in the single 
null coordinate system. The null dust solution is investigated in section III. 
Section IV is devoted to discuss the
high-speed approximation scheme for the general type I mater. The effects
of pressure on the high-speed gravitational collapse are discussed
in section V. Finally, the summary of the results is
given in section VI.

In this paper, we adopt the geometrized unit, i.e., $c=1=G$ and
follow the convention of the Riemann and metric tensors and the
abstract index notation adopted in the textbook by Wald
\cite{R25}; the latin indices denote the type of a tensor, whereas the 
Greek indices denote the components of a tensor. 

\section{Spherically symmetric matter with anisotropic pressure}

We focus on the spacetime with spherical symmetry. For later
convenience, we adopt the single null coordinate system in which the
line element is given by
\begin{equation}\label{E:1}
ds^2=-A(v,r)dv^2+2B(v,r)dvdr+r^2\left(d\theta^2+\sin^2\theta
d\phi^2\right),
\end{equation}
where we assume that $B$ is positive, and this assumption implies
that the coordinate $v$ is the advanced time, i.e., a radial curve
of constant $v$ is future directed ingoing null. Then the non-trivial components of
the Einstein equations are given by
\begin{eqnarray}\label{E:2}
\frac{A}{rB^2}\left(\frac{A'}{A}-\frac{2B'}{B}\right)
-\frac{1}{r^2}\left(1-\frac{A}{B^2}\right)&=&8\pi T^v{}_v,
\\\label{E:3}
\frac{2B'}{rB^2}&=&8\pi T^v{}_r, \\\label{E:4}
\frac{A}{rB^2}\left(\frac{2\dot{B}}{B}-\frac{\dot{A}}{A}\right)
&=&8\pi T^r{}_v, \\\label{E:5}
\frac{A'}{rB^2}-\frac{1}{r^2}\left(1-\frac{A}{B^2}\right) &=&8\pi
T^r{}_r, \\\label{E:6}
\frac{1}{B^2}\left(\dot{B}'+\frac{A''}{2}\right)
-\frac{B'}{B^3}\left(\dot{B}+\frac{A'}{2}\right)
+\frac{A}{rB^2}\left(\frac{A'}{A}-\frac{B'}{B}\right) &=&8\pi
T^\theta{}_\theta,\label{E:7}
\end{eqnarray}
where the dot denotes the derivative with respect to the advanced time $v$, while
the prime denotes the derivative with respect to the radial coordinate $r$.

We study the dynamics of the spherically symmetric type I matter 
whose stress-energy tensor is\cite{HE}
\begin{equation}\label{E:8}
T^a{}_b=\rho u^au_b+p_{\rm R}s^as_b+p_{\rm T}\Omega^a{}_b,
\end{equation}
where $u^a$ is 4-velocity of a constituent particle, 
$s^a$ is the unit radial vector normal to $u^a$, and $\Omega^a{}_b$ is 
defined by
\begin{equation}
\Omega^a{}_b=\delta^a{}_b+u^au_b-s^as_b,
\end{equation}
and thus $\rho$, $p_{\rm R}$ and $p_{\rm T}$ are the energy density, 
radial pressure and the tangential pressure, respectively. 

We write the components of $u^a$ and $s_a$ in the forms
\begin{eqnarray}
u^\mu&=&N\left(V,-1+V,0,0\right), \label{E:9} \\
s_\mu&=&NB\left(1-V,V,0,0\right), \label{E:9s}
\end{eqnarray}
where 
\begin{equation}\label{E:10}
N=\frac{1}{\sqrt{V\left\{2B+V(A-2B)\right\}}}.
\end{equation}
We define new variables $D$, $P_{\rm R}$ and $P_{\rm T}$ as
\begin{eqnarray}\label{E:12}
D&:=&\frac{N^2\sqrt{-g}(\rho+p_{\rm R})}{\sin\theta}
=\frac{r^2B(\rho+p_{\rm R})}{V\left\{2B+V(A-2B)\right\}}, \\\label{E:13}
P_{\rm R}&:=&\frac{N^2\sqrt{-g}p_{\rm R}}{\sin\theta}
=\frac{r^2Bp_{\rm R}}{V\left\{2B+V(A-2B)\right\}}, \\
P_{\rm T}&:=&\frac{N^2\sqrt{-g}p_{\rm T}}{\sin\theta}
=\frac{r^2Bp_{\rm T}}{V\left\{2B+V(A-2B)\right\}},
\end{eqnarray}
where $g$ is the determinant of the metric tensor. Then the
stress-energy tensor is written in the form
\begin{equation}\label{E:14}
T^a{}_b=\frac{1}{r^2 B}\left[ Dk^ak_b
+V\left\{2B+V(A-2B)\right\}
\left\{P_{\rm R}\delta^a{}_b +(P_{\rm T}-P_{\rm R})\Omega^a{}_b\right\} 
\right],
\label{eq:st-tensor}
\end{equation}
where the components of the vector field $k^a$ are
\begin{equation}\label{E:15}
k^\mu=\frac{u^\mu}{N}=\left(V,-1+V,0,0\right).
\end{equation}

\section{Null dust limit}

\subsection{Metric}

It is easy to see that in the limit of $V\rightarrow0$ with $D$, 
$P_{\rm R}$ and $P_{\rm T}$ fixed, the stress-energy tensor (\ref{E:14}) becomes that
of the null dust which belongs to the type II matter\cite{HE}, 
\begin{equation}\label{E:16}
T^a{}_b=\frac{D}{r^2 B}k^a k_b,
\end{equation}
where in this limit we have 
\begin{equation}\label{E:17}
k^\mu=(0,-1,0,0).
\end{equation}
We can easily check that $B^{-1}k^\mu$ is the tangent of the ingoing
null geodesic, i.e.,
\begin{equation}\label{E:18}
k^a\nabla_a\left(B^{-1}k_b\right)=0.
\end{equation}
Thus the equation of motion $\nabla_aT^a{}_b=0$ leads
\begin{equation}\label{E:19}
\nabla_a\left(r^{-2}Dk^a\right)=0.
\end{equation}
The above equation reduces to
\begin{equation}\label{E:20}
\left(BD\right)'=0,
\end{equation}
and thus we find that $BD$ is a function of the advanced time
$v$ only. Further since Eq.(\ref{E:3}) reduces to $B'=0$, we have
$B=B(v)$. Therefore, we have $D=D(v)$. If we introduce a new
advanced time $\bar{v}$ defined by
\begin{equation}\label{E:21}
\bar{v}=\int B(v)dv,
\end{equation}
we have a new line element
\begin{equation}\label{E:22}
ds^2=-\bar{A}(\bar{v},r)d\bar{v}^2+2d\bar{v}dr
+r^2\left(d\theta^2+\sin^2\theta d\phi^2\right),
\end{equation}
where $\bar{A}=A/B^2$.  Thus without loss of generality, we can
assume $B=1$ and will do so below in this section. 

The remaining non-trivial components of the Einstein equations
become 
\begin{eqnarray}\label{E:23}
\frac{A'}{r}-\frac{1}{r^2}\left(1-A\right)&=&0,
\\\label{E:24}
\dot{A}&=&-\frac{8\pi D}{r}, \\\label{E:25}
\frac{A''}{2}+\frac{A'}{r}&=&0.
\end{eqnarray}
From Eq.(\ref{E:23}), we obtain
\begin{equation}\label{E:26}
A=1-\frac{2M(v)}{r}.
\end{equation}
From Eq.(\ref{E:24}) and the above equation, we have
\begin{equation}\label{E:27}
\frac{dM}{dv}=4\pi D(v).
\end{equation}
If we fix the dependence of $D$ on $v$, the solution is completely
determined. This solution is known as the Vaidya solution.

\subsection{Central singularity}

The precursory singularity in spherical gravitational collapse formed at the symmetry 
center can be naked singularity\cite{Lake92}.  
The central singularity in Vaidya spacetime can be naked if the mass 
function $M(v)$ satisfies some condition\cite{Hiscock_WE,Kuroda,Joshi_D}.
Here we give a brief review about this issue by following the analysis in Ref.\cite{Kuroda}. 

The future directed ingoing null condition is $v=$const, whereas 
the future directed outgoing null condition is given by 
\begin{equation}\label{null-condition}
\frac{dr}{dv}=\frac{1}{2}\left(1-\frac{2M}{r}\right).
\end{equation}
We assume that $D(v)$ has a compact support so that we have 
$M(v)=0$ for $v\leq0$ and $M(v)\neq0$ for $v>0$. This assumption guarantees that the 
symmetry center $r=0$ is regular initially, i.e., for $v\leq0$. 
The central singularity forms at $v=0=r$. 
To know whether the central singularity is naked, we 
investigate the existence of the outgoing null geodesics from $v=0=r$. 

Suppose that the $r=r_0(v)$ is a solution for Eq.(\ref{null-condition}) which 
emanates from the central singularity $v=0=r$. The mass 
function is written by using this solution as
\begin{equation}
M(v)=\frac{1}{2}r_0(v)\left[1-2\dot{r}_0(v)\right].
\end{equation}
The above equation is just the condition on the mass function for which 
the central singularity becomes naked. 
It should be noted that there is only one ingoing null geodesic $v=0$ 
that hits the central singularity $v=0=r$. Hence,
if there is a one-parameter family of null geodesics 
emanating from the central singularity, the central naked singularity is null,
while, if there is only one, the central singularity is instantaneous, i.e., 
an event in conformally extended spacetime manifold.  
Solutions in such a one-parameter family are written in the form
\begin{equation}\label{family}
r(v)=r_0(v)+z(v;\sigma),
\end{equation}
where $\sigma$ parametrizes the solutions.
Substituting the above form into 
Eq.(\ref{null-condition}), we have
\begin{equation}\label{family-eq}
\frac{dz}{dv}=-\frac{z(2\dot{r}_0-1)}{2(z+r_0)}.
\end{equation}
We search for the solutions which behave as $z\rightarrow0$ for 
$v\rightarrow0$. 

\begin{figure}
{\resizebox{10cm}{!}{\includegraphics{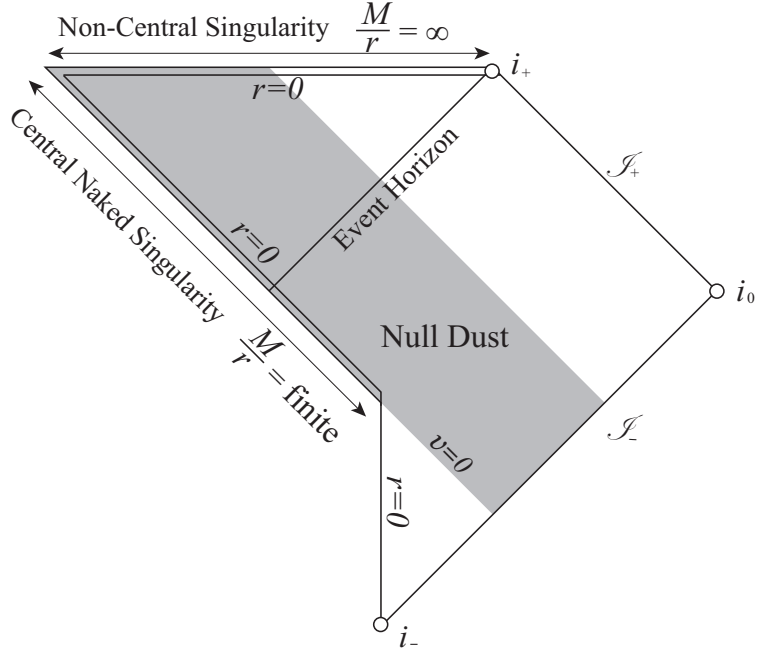}}}
\caption{The conformal diagram of the Vaidya spacetime in the cases (i) and (ii). 
The central naked singularity is null and thus there is a family of the
future directed outgoing radial null geodesics.
}
\label{fg:case-1}
\end{figure}

\begin{figure}
{\resizebox{5.5cm}{!}{\includegraphics{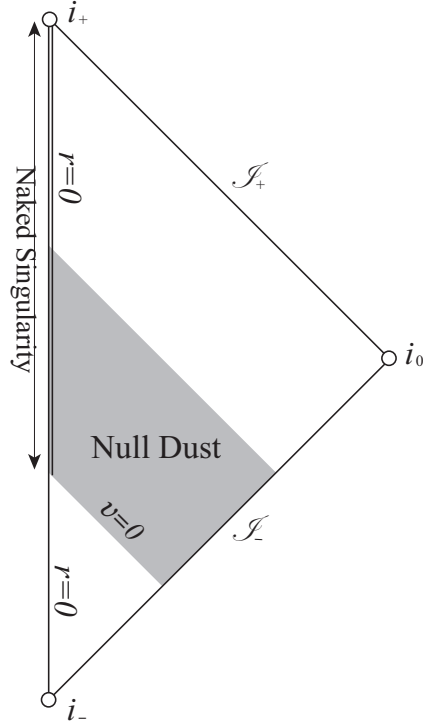}}}
\caption{The conformal diagram of the Vaidya spacetime in the case (iii). 
The central naked singularity is instantaneous. The final product of 
the gravitational collapse is the Schwarzschild spacetime with negative mass. 
}
\label{fg:case-2}
\end{figure}

\begin{figure}
{\resizebox{6.5cm}{!}{\includegraphics{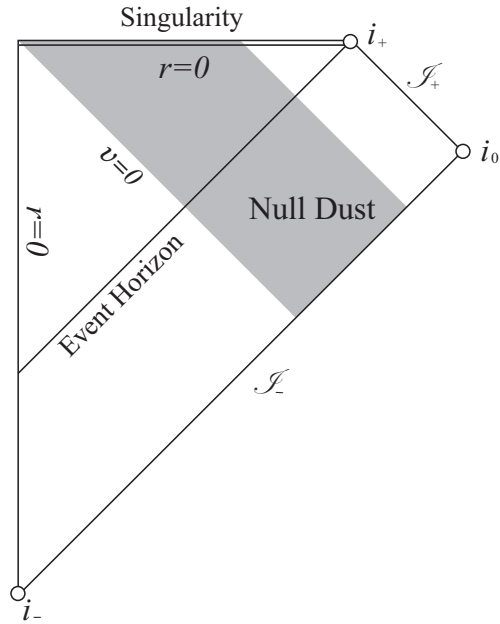}}}
\caption{The conformal diagram of the Vaidya spacetime in the case that 
the central singularity is not naked. The mass function $M$ 
behaves as $M\sim \mu v^\gamma$ with $\mu>1/16$ and $\gamma=1$ near $v=0$. 
}
\label{fg:case-3}
\end{figure}

Suppose $r_0\sim \beta v^\alpha$ near the central singularity, where both of $\alpha$ 
and $\beta$ are positive. Then we find following three cases: 

(i) $M(v) \sim \beta v^\alpha /2~~~~~~~~~~$ for $\alpha>1$; 

(ii) $M(v) \sim \beta(1/2-\beta) v~~$ for $\alpha=1$; 

(iii) $M(v)\sim -\alpha\beta^2 v^{2\alpha-1}~~~~$ for $\alpha<1$.

\noindent
For case (i), the solution for Eq.(\ref{family-eq}) is given by 
\begin{equation}
z\sim \sigma v^{-\alpha}\exp\left(-\frac{1}{2\beta(\alpha-1)v^{\alpha-1}}\right)
~~~~{\rm and}~~~~z\sim \frac{v}{2}.
\end{equation} 
For case (ii), we have, for $\beta\neq 1/4$, 
\begin{equation}\label{case2-sol}
v\sim \sigma|z|^{\frac{2\beta}{1-2\beta}}+\frac{2z}{1-4\beta},
\end{equation}
while we have, for $\beta=1/4$, 
\begin{equation}
v \sim \sigma z+4z\ln|z|.
\end{equation}
We see from Eq.(\ref{case2-sol}) that $\beta$ should be less than 1/2 so that 
$z\rightarrow0$ for $v\rightarrow0$. This condition guarantees that the mass 
function $M(v)$ is non-negative near $v=0$. 
For case (iii), the mass function $M(v)$ is negative near $v=0$. Thus in this case, 
the spacetime singularity of $v>0$ and 
$r=0$ is necessarily naked but it is non-trivial whether 
the central singularity at $v=0=r$ is null. We can see that 
the solution for Eq.(\ref{family-eq}) is
only $z\sim -2\beta v^\alpha$, and from Eq.(\ref{family}), we have
\begin{equation}
r\sim -\beta v^\alpha <0.
\end{equation}
Since $r$ should be non-negative, 
this is an unacceptable solution. Thus for case (iii), 
the solution $r=r_0(v)$ is the unique outgoing radial null geodesic which emanates 
from the central singularity. 
Further the radial curve $v=0$ is the unique future directed ingoing radial null 
which hits the central singularity. Thus we may conclude that the central 
singularity of negative mass function is instantaneous. 

We see from the above results that if the mass function $M$ behaves as 
$M\sim \mu v^\gamma$  near $v=0$ with $\mu>1/16$ and $\gamma=1$, 
then the central singularity is not naked. 

We depict the conformal diagrams 
for the cases (i) and (ii) in Fig.1, while for the case (iii) in 
Fig.2. The covered case is depicted in Fig.3.

\section{High-speed approximation scheme}

The high-speed approximation is a kind of linear perturbation scheme
in which $V$ is a small variable of order $\epsilon$. Then we
write
\begin{eqnarray}\label{E:28}
A&=&1-\frac{2M_B(v)}{r}+\delta_A(v,r), \\\label{E:29}
B&=&1+\delta_B(v,r), \\\label{E:30}
D&=&D_B(v)[1+\delta_D(v,r)],
\end{eqnarray}
where $\delta_A$, $\delta_B$ and $\delta_D$ are also assumed to be
$O(\epsilon)$, and
\begin{equation}\label{E:31}
\dot{M}_B=4\pi D_B.
\end{equation}
We take into account the terms up to the first order of
$\epsilon$. Hereafter, we assume that $V$ is non-negative so that 
$u^a$ is a causal vector.  As in the previous section, we also assume here 
that $D_B(v)$ has a compact support so that we have $M_B(v)=0$ for 
$v\leq0$ and $M_B(v)\neq0$ for $v>0$. 

From Eq.(\ref{E:14}), we have
\begin{equation}\label{E:32}
T^v{}_r=\frac{D}{r^2}V^2.
\end{equation}
This equation and Eq.(\ref{E:3}) lead $B'={\cal O}(\epsilon^2)$,
and thus we have
\begin{equation}\label{E:33}
B=1+\delta_B(v)+{\cal O}(\epsilon^2) .
\end{equation}
The above result means that $B$ is the function of only $v$ up to
the first order of $\epsilon$, and hence, without loss of
generality, we can assume $B=1$ up to this order 
by the same reason as in the case of the Vaidya solution.
Then Eqs.(\ref{E:2}) and (\ref{E:5}) agree with each other up to
the first order of $\epsilon$. 

The remaining apparently
independent components of the Einstein equations of $O(\epsilon)$ are
\begin{eqnarray}\label{E:34}
(r\delta_A)'&=&-8\pi(D_B-2P_{\rm R})V,
\\\label{E:35}
r\dot{\delta}_A&=&8\pi D_B
\left[\left(1+\frac{2M_B}{r}\right)V-\delta_D\right],
\\\label{E:36} (r\delta_A)''&=&\frac{32\pi}{r} P_{\rm T}V.
\end{eqnarray}
The equations of motion for the matter $\nabla_aT^a{}_b=0$ of
order $O(\epsilon)$ are given by
\begin{eqnarray}\label{E:37}
\left[(D_B-2P_{\rm R})V\right]\dot{}
+D_B\left[\left(1+\frac{2M_B}{r}\right)V-\delta_D\right]'&=&0,
\\\label{E:38}
\left[(D_B-2P_{\rm R})V\right]'+\frac{4}{r}P_{\rm T}V&=&0.
\end{eqnarray}
Eq.(\ref{E:36}) is not necessary, since this equation is derived
from Eqs.(\ref{E:34}) and (\ref{E:38}). Differentiating
Eq.(\ref{E:34}) with respect to $r$, we have
\begin{equation}\label{E:39}
(r\delta_A)''=-8\pi\left[(D_B-2P_{\rm R})V\right]'.
\end{equation}
Substituting Eq.(\ref{E:38}) into the right hand side of the above
equation, we have Eq.(\ref{E:36}). Eq.(\ref{E:37}) is 
also not necessary, since this equation is derived from
Eqs.(\ref{E:34}) and (\ref{E:35}). Differentiating Eq.(\ref{E:34})
with respect to $v$, we have
\begin{equation}\label{E:40}
(r\dot{\delta}_A)'=-8\pi\left[(D_B-2P_{\rm R})V\right]\dot{}~.
\label{eq:vv-del}
\end{equation}
Differentiating Eq.(\ref{E:35}) with respect to $r$, we have
\begin{equation}\label{E:41}
(r\dot{\delta}_A)'=8\pi D_B
\left[\left(1+\frac{2M_B}{r}\right)V-\delta_D\right]'.
\label{eq:rv-del}
\end{equation}
From Eqs.(\ref{E:40}) and (\ref{E:41}), we have Eq.(\ref{E:37}).
Thus independent equations are only Eqs.(\ref{E:34}),
(\ref{E:35}) and (\ref{E:38}).

\section{The analysis of the high-speed collapse}

In order to write down formal solutions for the perturbation variables, 
we introduce new variables $k(v,r)$ and $w(v,r)$, defined by
\begin{equation}\label{E:42}
p_{\rm R}=k(v,r)\rho~~~~{\rm and}~~~~p_{\rm T}=w(v,r)\rho. \label{eq:EOS}
\end{equation}
Here, we briefly review the so-called weak, strong and dominant energy 
conditions, which might be satisfied by physically reasonable matter. 
These conditions are rewritten in the form of the conditions on the 
values of $k$ and $w$. 
The weak energy condition leads the following conditions on $\rho$, 
$p_{\rm R}$ and $p_{\rm T}$, 
\begin{equation}
\rho\geq0,~~~~\rho+p_{\rm R}\geq0~~~~ {\rm and}~~~~\rho+p_{\rm T}\geq0;
\end{equation}
the strong energy condition adds one more condition
\begin{equation}
\rho+p_{\rm R}+2p_{\rm T}\geq 0;
\end{equation}
the dominant energy condition leads further two conditions, 
\begin{equation}
\rho\geq|p_{\rm R}|~~~~{\rm and}~~~~\rho\geq|p_{\rm T}|.
\end{equation}
Assuming that $\rho$ is non-negative,
all the three energy conditions are guaranteed, if and only if following conditions 
for $k$ and $w$ are satisfied:
\begin{equation}
  -1\leq k \leq 1, ~~~~-1\leq w \leq 1~~~~{\rm and}~~~~1+k+2w\geq0. \label{eq:energy}
\end{equation}
Hereafter we assume the non-negativity of $\rho$. 
It is worthy to notice that both of $k$ and $w$ are bounded 
above and below by virtue of the energy conditions. 

From Eq.(\ref{eq:EOS}), we have
\begin{equation}\label{E:43}
\frac{P_{\rm R}}{D}=\frac{p_{\rm R}}{\rho+p_{\rm R}}=\frac{k}{k+1} ~~~~~{\rm and}~~~~~
\frac{P_{\rm T}}{D}=\frac{p_{\rm T}}{\rho+p_{\rm R}}=\frac{w}{k+1}.
\end{equation}
Using the above equation, Eq.(\ref{E:38}) becomes
\begin{equation}\label{E:44}
\left(\frac{1-k}{1+k}V\right)' +\frac{4w}{r(1+k)}V=0,
\end{equation}
where we have used the fact $D_B=D_B(v)$. 
From the above equation, we have
\begin{equation}\label{E:45}
\left[\ln\Biggl|\frac{1-k}{1+k}V\Biggr|\right]'
=-\frac{4w}{r(1-k)}.
\end{equation}
Clearly, the high-speed approximation scheme is not applicable to the 
case of $k=\pm1$: the upper sign corresponds to the stiff matter 
and the lower sign corresponds to the cosmological-constant-like matter. 
Hereafter we assume $k\neq \pm1$. 

The formal solution of Eq.(\ref{E:45}) is given by
\begin{equation}
\frac{1-k}{1+k}V=C(v)\exp\left(-\int^r\frac{4w(v,x)}{1-k(v,x)}\frac{dx}{x}\right),
\label{formal-V-sol}
\end{equation}
where $C(v)$ is an arbitrary function of $v$ but it is set 
so that $V$ is positive at least initially.
Substituting the above equation into Eq.(\ref{E:34}) and integrating it, we have
\begin{equation}
\delta_A=-\frac{8\pi CD_B}{r}
\int^r
\exp\left(-\int^y\frac{4w(v,x)}{1-k(v,x)}\frac{dx}{x}\right)dy,
\label{formal-dA-sol}
\end{equation}
where we set the integration constant so that $\delta_A$ is finite in the limit 
of $r\rightarrow0$ if possible.
Then substituting Eqs.(\ref{formal-V-sol}) and (\ref{formal-dA-sol}) into Eq.(\ref{E:35}), 
we have
\begin{eqnarray}
D_B\delta_D&=&CD_B\left(\frac{1+k}{1-k}\right)\left(1+\frac{2M_B}{r}\right)
\exp\left(-\int^r\frac{4w(v,x)}{1-k(v,x)}\frac{dx}{x}\right)
\nonumber \\
&+&
\int^r
\left[(CD_B)\dot{~}
-CD_B\int^y\left(
\frac{4\dot{w}(v,z)}{1-k(v,z)}
+\frac{4w(v,z)\dot{k}(v,z)}{[1-k(v,z)]^2}
\right)\frac{dz}{z}\right]\nonumber \\
&\times&\exp\left(-\int^y\frac{4w(v,x)}{1-k(v,x)}\frac{dx}{x}\right)dy.
\label{formal-dD-sol}
\end{eqnarray}
Once $k$ and $w$ are determined, we can know the behavior of the first order perturbations 
by performing the integrations in the formal solutions (\ref{formal-V-sol}), 
(\ref{formal-dA-sol}) and (\ref{formal-dD-sol}). 

The behavior of the solutions (\ref{formal-V-sol})--(\ref{formal-dD-sol}) 
near the spacetime singularity will be determined by the asymptotic values of 
$k(v,r)$ and $w(v,r)$ in the limit that the hollow sphere shrinks to its symmetry 
center. 
Thus the solutions obtained by assuming the constancy of $k$ and $w$ will 
give us sufficient information about what we would like to know. 
Hereafter, we assume that $k$ and $w$ are constant.
Integration in Eq.(\ref{formal-V-sol}) is then easily performed and we obtain 
\begin{eqnarray}
V&=&\frac{1+k}{1-k}C(v)r^{-\frac{4w}{1-k}}.  \label{const-V-sol}
\end{eqnarray}
Using the above result, integrations in Eqs.(\ref{formal-dA-sol}) and 
(\ref{formal-dD-sol}) are also easily performed, and we obtain 
\begin{eqnarray}
\delta_A&=&-8\pi CD_B F(r;k), \label{const-dA-sol} \\
D_B \delta_D&=&CD_B\left(\frac{1+k}{1-k}\right)\left(1+\frac{2M_B}{r}\right)
r^{-\frac{4w}{1-k}}+(\dot{D}_BC+D_B\dot{C}) F(r;k), \label{const-dD-sol}
\end{eqnarray}
where
\begin{equation}
F(r;k)=
\left\{
\begin{array}{ll}
(1-k-4w)^{-1}(1-k)r^{-\frac{4w}{1-k}}
& {\rm for}~~1-k-4w\neq0, \\
r^{-1}\ln r & {\rm for}~~1-k-4w=0.
\end{array}
\right.
\end{equation}

From the causality requirement\cite{EMM}, we assume $k<1$ and $w\leq 1$. Then 
we study the following three cases, 
$0<w\leq1$, $-(1-k)/4<w\leq 0$ and $w\leq-(1-k)/4$, separately, below.  

\subsection{$0<w\leq1$}

It can be seen from Eq.(\ref{const-V-sol})  
that $V$ diverges in the limit of $r\rightarrow0$. 
This implies that, in this case, the high-speed collapse is necessarily decelerated 
by the pressure effect so significantly that the high-speed approximation breaks down 
before the singularity formation. This is a somewhat unexpected result, since, 
at first glance, one might infer that the pressure effect would be negligible in the 
situation with the strong gravity near the spacetime singularity.

\subsection{$-(1-k)/4<w\leq 0$}\label{sec:B}

It is easily seen from Eqs.(\ref{const-V-sol}) and (\ref{const-dA-sol}) 
that both $V$ and $\delta_A$ are finite in the limit of $r\rightarrow0$. 
Eq.(\ref{const-dD-sol}) leads 
\begin{equation}
D_B\delta_D \sim \frac{2M_B}{r}CD_Br^{-\frac{4w}{1-k}}~~~~~{\rm near}~~r=0.  
\label{eq:delD-0}
\end{equation}
Here note that by assumption, we have
\begin{equation}
0\leq-\frac{4w}{1-k}<1. \label{w-inequality}
\end{equation}
$M_B/r$ diverges in the limit of $r\rightarrow0$ with $v$ fixed in the domain 
with non-vanishing $M_B$. Thus, due to Eq.(\ref{w-inequality}), 
the density perturbation $D_B\delta_D$ also diverges in the limit of $r\rightarrow0$ 
within this domain, if it initially does not 
vanish there. The high-speed approximation in this domain breaks down  
when the matter particles approach the symmetry center $r=0$. 

Since $M_B$ vanishes at $v=0$, the behavior of the density perturbation $D_B\delta_D$ 
in the neighborhood of the central singularity might be  
different from the above case.  
The behavior of $M_B/r$ near the central singularity has already been shown in 
Sec.III. Our special interests are in the cases (i) and (ii), since the 
the background central singularity has the extent in the future directed ingoing 
null direction, while 
in the case (iii) and in the covered case, the central singularity is 
instantaneous (see Figs.1-3). Thus we focus on the cases (i) and (ii) in which 
the mass function $M_B$ behaves as $M_B\sim \mu v^\alpha$ with $\alpha\geq 1$. 
In these cases, $M_B/r$ is finite at the central singularity 
in the background Vaidya spacetime. 
Thus the central singularity will be very similar to 
the central naked singularity of the Vaidya solution. 

Here it should be noted that this case includes a case of dust $k=w=0$. 
The dust is described by the Lema\^itre-Tolman-Bondi solution and it is well known 
the central singularity can also be naked\cite{ES,Chris,Newman,JD93}. 
The present result implies that the central singularity formed by the 
gravitational collapse of a hollow dust sphere is 
very similar to that formed by the null dust, whereas the non-central singularity 
is not so. The latter is in contrast to the cylindrically symmetric case in which 
the cylindrical hollow dust collapsing with very high speed is well described 
by the null dust solution even at the spacetime singularity\cite{R22}. 

\subsection{$w\leq-(1-k)/4$}

In contrast to the case {\bf B}, the perturbation variables 
$V$ and $\delta_A$ vanish in the limit of 
$r\rightarrow0$ with $v$ fixed. 
In the case of $w<-(1-k)/4$, $D_B\delta_D$ also vanishes in the same limit,  
whereas it is finite in the case of $w=-(1-k)/4$. 
Thus the spacetime structure in the neighborhood 
of the singularity at $r=0$ is very similar to the Vaidya solution. 
Even if the radial pressure is positive, the high-speed 
approximation is consistent until the spacetime singularity forms. The consistency 
of the high-speed approximation depends not on the radial pressure 
but on the tangential one. Here note that all of the energy 
conditions Eq.(\ref{eq:energy}) hold only if 
\begin{equation}
k\geq-\frac{1}{3}
\end{equation}
holds. 

Recent observations imply the acceleration of cosmic volume 
expansion\cite{Reiss,Perl,CMB}, and this 
means the existence of unknown matter components with violation of the 
strong energy condition. 
Thus it might be important to consider cases with the violation of the 
energy conditions as special examples of case {\bf C}. 
A case of $k$ smaller than $-1/3$ corresponds to the so-called dark energy, and  
thus the result obtained here implies that 
the spacetime singularity formed by spherically symmetric high-speed collapse of 
the dark energy is  similar to that of the Vaidya solution. 
Further, it is worthy to note that in the phantom energy case $k<-1$
which is the special case of the dark energy\cite{phantom},  
$D_B$ is negative and thus the mass function $M_B$ is also 
negative by Eq.(\ref{E:31}). Hence the first order solution 
of $k<-1$ implies that the high-speed collapse of the phantom 
energy forms the timelike singularity similar to that 
in the Schwarzschild spacetime with a negative mass (see Fig.2). 

\section{Summary and Discussion}

Gravitational collapse is one of the most important topics in 
gravitational physics. The cosmic censorship conjecture 
provides major motivation to study this issue. 
Since there is no theorem proving or disproving this conjecture 
or no theorem stating the generic feature of physical spacetime singularities, 
it is interesting to investigate this issue 
in the situation different from previously studied ones. 

This paper continues to study this issue and provides an extension
of the previous work on high-speed cylindrical collapse of perfect
fluid \cite{R23} to spherically symmetric spacetime with type I matter.  To see the
pressure effects on high-speed approximation scheme, assuming that 
the energy density $\rho$ is non-negative, we have
studied a linear equation of state for the radial and tangential pressures, 
i.e., $p_{\rm R}=k\rho$ and 
$p_{\rm T}=w\rho$ with constant $k$ and $w$, in detail. 
By the causality requirement, we have restricted our attention to the case of $k<1$ 
and $w\leq1$. (The causality requirement implies 
$k\leq1$. The reason why $k=1$ is excluded is that the high-speed approximation 
is not applicable to the cases of $k=\pm1$.) 

One might think that the collapsing speed becomes very large 
due to the strong gravity just before the formation of the spacetime singularity. 
However, we have found that, in the case of positive tangential pressure $0<w\leq 1$, 
the large initial imploding velocity is necessarily decelerated by the pressure effect 
and thus the high-speed approximation 
scheme becomes invalid before the singularity formation.

In the case of the negative or vanishing tangential pressure $w\leq0$, 
the behaviors of the perturbation variables are different from the case of 
$0<w\leq1$. In the case of $-(1-k)/4<w\leq0$, all of the perturbation are 
finite at the central singularity, 
if the background central singularity is null and naked, 
although the density perturbation blows up at the non-central singularity 
of the background Vaidya spacetime. 
This result implies that the central singularity will be null and naked 
like as that of the Vaidya spacetime, for the case of $-(1-k)/4<w\leq0$. 
In the cases of $w\leq-(1-k)/4$, all of the perturbation variables are finite 
everywhere. This result implies that, in this case, the geometrical structure 
near not only the central singularity but also the non-central singularity 
is very similar to that of the Vaidya solution. 
Here it is worthy to note that as long as $k>-1/3$, 
this case satisfies all of the physically reasonable energy conditions, i.e., 
the weak, strong and dominant energy conditions. 
The result obtained here strongly suggests that the spacetime structure realized by 
the spherical matter with large enough tangential 
tension is well described by the Vaidya solution even if 
the physically reasonable energy conditions are satisfied. 

It is a remarkable result that the consistency 
of the high-speed approximation depends not on the radial pressure $p_{\rm R}$ 
but on the tangential one $p_{\rm T}$. 
In the case of $k\leq 0$, the gradient of radial pressure will not 
stop the high-speed collapse. Thus it is reasonable that the consistency of the high-speed 
approximation depends on the only tangential pressure. 
It is non-trivial that even if $k>0$, the consistency of the high-speed approximation 
also depends on the only tangential pressure. At first glance, the gradient of the radial 
pressure with $k>0$ seems to affect the gravitational collapse, 
but this is not true. The reason is that if the sound speed in the radial direction 
is less than the speed of light ($0<k<1$), the spacetime singularity formation by the 
nearly light-speed collapse can be completed before the effect of 
radial pressure gradient spreads out, since the sound cone is significantly 
narrowed down in the frame in which 
the matter moves with nearly the speed of light. This is also the reason why the 
high-speed approximation is not applicable to the case of $k=1$. In this case,  
the sound speed in the radial direction is equal to the speed of light 
and thus the sound cone in the radial direction is equivalent to the light cone which is 
Lorentz invariant.  As a result, the effect of the radial pressure gradient can 
spread out before the singularity formation by the nearly light-speed collapse is completed. 
The perturbative construction of approximate solutions 
on the background null dust solution is impossible in the case of $k=1$. 

The dark energy case, $k\leq-1/3$ and $w<-(1-k)/4$, might be important in connection 
to the issue of the accelerated cosmic expansion\cite{Reiss,Perl,CMB},
although the strong energy is not satisfied in this case. 
These are included in the case of $w<-(1-k)/4$. 
The repulsive gravity due to 
the dark energy cannot decelerate the high-speed collapse, and 
the formed central singularity can be null, and the non-central 
part of the spacetime singularity will be spacelike. 
As mentioned in the above, the pressure gradient
force of the dark energy will also not stop the collapse 
but rather accelerate it.
This property of the dark energy will be the reason 
why the high-speed collapse becomes a good approximation at the singularity 
formation in the present case and also 
why black holes may grow self-similarly due to the accretion of 
dark energy in the accelerated universe~\cite{hmc2008,mhc2008}.

Finally, it should be noted that the present results are valid up to the only first order 
and could be modified by the higher order effects. 
Thus we need to investigate the higher order, but this is a future work. 

\section*{Acknowledgments}

MS and ZA would like to thank the Higher Education Commission Islamabad,
for its financial support through the {\it Indigenous PhD 5000
Fellowship Program Batch-I}. 
TH was supported by 
the Grant-in-Aid for Scientific
Research Fund (Young Scientists (B) 18740144) of the Ministry of Education, 
Culture, Sports, Science and Technology of Japan.
KN is grateful to colleagues in the astrophysics and 
gravity group of Osaka City University for helpful discussion and criticism.


\vspace{0.5cm}

\begin{thebibliography}{99}

\bibitem{R1} P.S. Joshi, {\it Global Aspects in Gravitation and
             Cosmology} (Oxford University Press, 1993).

\bibitem{R2} R.~Penrose, Phys. Rev. Lett. {\bf 14}, 57 (1965);\\
             S.W. Hawking, {\it Proc.R. Soc. London} {\bf A300}, 187 (1967);
             S.W. Hawking and R. Penrose, {\it Proc. R. Soc. London} {\bf
             A314}, 529 (1970).

\bibitem{R3} R.~Penrose, Nuovo Cimento {\bf 1}, 252 (1969); in {\it General Relativity, an 
    Einstein Century Survey}, edted {\bf [edited]} by S.W.~Hawking and W.~Israel 
    (Cambridge University Press, Cambridge, England, 1979), p.581.

\bibitem{R3-1} L.M. Burko, Phys. Rev. Lett. {\bf 79}, 4985 (1997).

\bibitem{R3-2} M. Dafermos, arXiv: gr-qc/0401121.

\bibitem{R4} J.R. Oppenheimer and H. Snyder, Phys. Rev. {\bf 56}, 455 (1939).

\bibitem{R5} R. Goswami and P.S. Joshi, Phys. Rev. {\bf D69}, 027502 (2004).

\bibitem{R6} R. Goswami, P.S. Joshi, C. Vaz and L. Witten, Phys.
             Rev. {\bf D70}, 084038 (2004).

\bibitem{R7} K. Nakao, H. Iguchi and T. Harada, Phys. Rev. {\bf D63}, 084003 (2001).

\bibitem{R8} T. Harada, H. Iguchi and K. Nakao, Prog. of Theor. Phys. {\bf
             107}, 449 (2002).

\bibitem{R9} H. Iguchi, K. Nakao and T. Harada, Phys. Rev. {\bf
             D57}, 7262 (1998).

\bibitem{R10} P.S. Joshi, N. Dadhich and R. Maartens, Phys. Rev. {\bf
              D65}, 101501 (2002).

\bibitem{R11} P.S. Joshi and T.P. Singh, Phys. Rev. {\bf
              D51}, 6778 (1995).

\bibitem{R12} T.P. Singh, Class. Quantum Grav. {\bf 16}, 3307 (1999).

\bibitem{R13} A. Ori and T. Piran, Phys. Rev. Lett. {\bf 59}, 2137 (1987).

\bibitem{R14} A. Ori and T. Piran, Gen. Relativ. Grav. {\bf 20}, 7 (1988).

\bibitem{R15} A. Ori and T. Piran, Phys. Rev. {\bf D42}, 1068 (1990).

\bibitem{R16} T. Foglizzo and R.N. Henriksen, Phys. Rev. {\bf D48}, 4645 (1993).

\bibitem{R17} T. Harada, Phys. Rev. {\bf D58}, 104015 (1998).

\bibitem{R18} T. Harada and H. Maeda, Phys. Rev. {\bf D63}, 084022 (2001).

\bibitem{R19} T. Harada, Class. Quantum Grav. {\bf 18}, 4549 (2001).

\bibitem{R20} R. Giambo, F. Giannoni, G. Magli and P. Piccione,
              Class. Quantum Grav. {\bf 20}, 4943 (2003).

\bibitem{R21} R. Goswami and P.S. Joshi, Class. Quantum Grav. {\bf 21}, 3645 (2004).

\bibitem{R22} K. Nakao and Y. Morisawa, Class. Quantum Grav. {\bf 21}, 2101 (2004).

\bibitem{R23} K. Nakao and Y. Morisawa, Prog. Theor. Phys. {\bf 113}, 73 (2005).

\bibitem{R24} M. Sharif and Z. Ahmad, Gen. Relativ. Grav. {\bf 39}, 1331 (2007).

\bibitem{HE} S.W. Hawking and G.F.R. Ellis, {\it The large scale structure of space-time} 
(Cambridge University Press, 1973) p.89.

\bibitem{DJ1994} I.H. Dwivedi and P.S. Joshi, Commun. Math. Phys. {\bf 166}, 117 (1994).

\bibitem{GJ2007} R. Goswami and P.S. Joshi, Phys. Rev. D {\bf 76}, 084026 (2007).

\bibitem{Magli1998}
G. Magli,  Class. Quant. Grav. {\bf 14} (1997) 1937; Class. Quant. Grav. {\bf 15} 3215 (1998).

\bibitem{MGJM2002}
S. M. C. V. Goncalves, S. Jhingan, G. Magli, Phys.Rev. {\bf D65} 064011 (2002). 

\bibitem{TIN1998}
T.Harada, H. Iguchi and K.Nakao, Phys. Rev. D {\bf 58} 041502 (1998).  

\bibitem{TNI1999}
T.Harada, K. Nakao and H. Iguchi, Class. Quantum Grav. {\bf 16} 2785 (1999).  

\bibitem{GJ2002}
R. Goswami and P. S. Joshi, Class. Quantum Grav.{\bf 19}, 5229 (2002).

\bibitem{MHJN2007}
A.Mahajan, T. Harada, P.S. Joshi and K. Nakao, Prog. Theor. Phys. {\bf 118} 865 (2007). 

\bibitem{R25} R.M. Wald, {\it General Relativity} (The University of Chicago Press, 1984).

\bibitem{Lake92} K. Lake, Phys. Rev. Lett. {\bf 68}, 3129 (1992).

\bibitem{Hiscock_WE} W.A. Hiscock, L.G. Williams and D.M. Eardley, 
Phys. Rev. D {\bf 26}, 751 (1982).

\bibitem{Kuroda} Y. Kuroda, Prog. Ther. Phys. {\bf 72}, 63 (1984).

\bibitem{EMM} G.~Ellis, R. Maartens and M. MacCallum, Gen. Rel. Grav. {\bf 39}, 1651 (2007)

\bibitem{Joshi_D} P.S. Joshi and I.H. Dwivedi, Phys. Rev. D {\bf 45}, 2147 (1992).

\bibitem{ES} D.M. Eardley and L. Smarr, Phys. Rev. D {\bf 19}, 2239 (1979).

\bibitem{Chris} D. Christodoulou, Commun. Math. Phys. {\bf 146}, 171 (1984).

\bibitem{Newman} R.P.A.C. Newman, Class. Quantum Grav. {\bf 3}, 527 (1986).

\bibitem{JD93} P.S. Joshi and I.H. Dwivedi, Phys. Rev. D {\bf 47}, 5357 (1993). 

\bibitem{Reiss} A.G. Reiss, et al., Astron. J. {\bf 116}, 1009 (1998). 

\bibitem{Perl} S. Perlmutter et al., Astrophys. J. {\bf 517}, 565 (1999). 

\bibitem{CMB} D.N. Spergel et al. (WMAP), Astrophys. J. Suppl. Ser. 
{\bf 148}, 175 (2003).

\bibitem{phantom} R.R. Caldwell, M. Kamionkowski and N.N. Weinberg, Phys. Rev. Lett. 
{\bf 91}, 071301 (2003).

\bibitem{hmc2008}
T. Harada, H. Maeda and B. J. Carr, Phys. Rev. D{\bf 77}, 
024022 (2008).

\bibitem{mhc2008}
H. Maeda, T. Harada and B. J. Carr, Phys. Rev. D{\bf 77}, 
024023 (2008).

\end{thebibliography}
\end{document}